
\raggedbottom
\hoffset=0cm\voffset=0cm
\magnification=\magstep1
\baselineskip=15pt

\def\ref#1{\noindent\item{[#1.]}}
\def\spa#1{\noindent\item {#1}}
\vskip 9cm
\hfill {\tenrm NUS/HEP/94201}
\par\noindent
\hfill {\tenrm hep-th/9407142}
\vskip 2cm
\centerline {\bf Generalized $q$-Oscillators and their Hopf Structures}
\vskip .7cm
{\centerline {C.H. Oh ${}^{\dag}$ and K. Singh $^{\ddag}$}}
\vskip .5cm
{\sl {\centerline {Department of Physics}}}
{\sl {\centerline {Faculty of Science}}}
{\sl {\centerline {National University of Singapore}}}
{\sl {\centerline {Lower Kent Ridge, Singapore 0511}}}
{\sl {\centerline {Republic of Singapore}}}
\vskip 2cm
{\bf {\centerline {Abstract}}}
\vskip .5cm
\noindent
We study the relationships among the various forms of the $q$
oscillator algebra and consider the conditions under which it
supports a
Hopf structure. We also present a generalization of this algebra
together with its corresponding Hopf structure. Its multimode
extensions are also considered.
\vskip 4.5truecm
\centerline{To appear in J. Phys.A: Math. Gen.}
\vskip.3cm
\hrule width6cm
\vskip .3truecm
{\sevenrm \dag E-mail: PHYOHCH@NUSVM.NUS.SG
\vskip.1truecm
\ddag E-mail: PHYSINGH@NUSVM.NUS.SG}
\vfil\eject
\noindent
{\bf 1. Introduction}
\par
Quantum groups or more precisely the quantized universal enveloping
algebras ${\cal U}_q({\cal L})$ of Lie algebras $\cal L$ first
emerged as the basic algebraic structures in the study of the
quantum Yang-Baxter equations [1]. It was later shown by Drinfeld
[2] that these structures could be described by a general class of
associative algebras, called Hopf algebras, which are neither
commutative nor cocommutative. Essentially the non-cocommutativity
is achieved by introducing a free parameter $q$ which is usually
called the deformation parameter.
\par
Now one of the most well studied example is that of the quantum
group ${\cal U}_q({\rm su(2)})$ (or sometimes denoted as
${\rm su}_{q}(2)$) which was first considered by
Skylanin [3] and independently by Kulish and Reshetikhin [4].
Recently this algebra has been realized in terms of a $q$-analogue
of the bosonic creation and annhiliation operators [5,6].
Indeed, Macfarlane [5] introduced these $q$-oscillators $a,~a^+$
by considering their action on a Hilbert space with basis $\{\vert
n>\},$ $n=0,1,2,...$ defined by
$$a\vert 0>=0,~~~~~~~\vert n>=([n]!)^{-1/2}(a^+)^{n}\vert 0>,
\eqno(1)$$
where
$$[n]={{q^n-q^{-n}}\over {q-q^{-1}}} ~~~~~{\rm and}~~~~~
[n]!=[n][n-1][n-2]...[1].$$
Then by setting
$$\eqalignno{
a^+a &= [N], &(2a)\cr
aa^+ &= [N+1], &(2b)\cr}$$
where $N$ satisfies
$$N\vert n>=n\vert n>\eqno(3a)$$
he was able to furnish a representation for the $q$-oscillators:
$$\eqalignno{a^+\vert n> &= [n+1]^{1/2}\vert n+1>, &(3b)\cr
 a\vert n> &= [n]^{1/2}\vert n-1>. &(3c)\cr}$$
Moreover in this representation, one also has the following
relations:
$$\eqalignno{
&aa^+ -qa^+a = q^{-N}, &(4)\cr
&aa^+ -q^{-1}a^+a = q^N, &(5)\cr}$$
besides
$$[N,a^+]=a^+,~~~~~~[N,a]=-a. \eqno(6)$$
Biedenharn [6] also independently arrived at similar results but
instead of starting with relations (2), he postulated (4) and (6)
with $q$ replaced by $q^{1/2}$.
\par
By using the Jordan-Schwinger construction, they gave a bosonic
realization of ${\rm su}_{q}(2)$.
Conversely, the $q$-oscillators can also be
obtained directly from the usual representation of ${\rm su}_{q}(2)$.
Ng [7] showed that by setting $j\to \infty,~m\to \infty$, in the
basis vectors spanning the Hilbert space of ${\rm su}_{q}(2)$,
the $q$-oscillators can be obtained which
satisfy relations (2).
\par
Although the $q$-oscillators have been primarily used in giving
realizations of  quantum groups, it itself may support a quantum
group structure. Indeed, Hong Yan [8] showed that the $q$-oscillator
algebra when expressed in a symmetric form could be endowed with a
non-cocommutative Hopf
structure. Instead of relations (2), he considered the the commutator
\footnote\dag {In ref.[8] the right hand side is expressed as
$[N+{{1}\over{2}}]-[N-{{1}\over{2}}]$.
Here we have made the replacement $N\to N+{{1}\over{2}}$,
to be consistent with the notation that we have adopted in this paper.}
$$[a,a^+]=[N+1]-[N].\eqno(7)$$
Then together with relations (6) he was able to construct a
non-trivial Hopf algebra.
It is worth noting that while relations (2) imply relation (7), the
converse is not true; it is only in the representation (3)
that the two are equivalent. In fact the same holds for relations
(4) and (5) in regard to (2) or (7).
\par
In the following section, we discuss some issues pertaining to this
inequivalence. In particular, we study the relationships among the
various forms of the $q$-oscillators. Moreover, we
also clarify some misleading notions in the literature about the
$q$-oscillator algebra when regarded as a quantum group. In section
3 we present a generalized deformed oscillator algebra which also
has a Hopf structure. Here, Hong Yan's algebra is recoverd as a special
case. The representation of this generalized algebra is also furnished.
In section 4 we consider its multimode extensions. Besides a set
consisting of mutually commuting oscillators, we also present a
multidimensional quantum group based on our generalization.
\vskip .5cm
\par \noindent
{\bf 2. $q$-Oscillator Algebras}
\par
The $q$-oscillator algebra consists of three elements $a$, $a^+$ and
$N$ defined by (6) together with {\it one} of the relations (2),
(4), (5), (7). In the following we will examine how the various forms
namely,
(2), (4), (5), and (7) of the oscillator algebra are related to
each other. Here relations (6) will be implicitly assumed as part
of the algebra. For clarity, we consider two at a time.
\vskip 1cm
\par \noindent
$\underline {\rm Case~(i):~Between~(2)~and~(4).}$
\par \noindent
Starting with (2), we show that it implies (4). Indeed, by
substituting (2) into the LHS of (4), we have
$$aa^+-qa^+a=[N+1]-q[N]=q^{-N}.\eqno(8)$$
which is precisely the RHS of (4). On the other hand to see
whether (4) implies (2) we construct the Casmir operator
for the algebra defined by (4) and (6). Now,
it is easy to verify that
$${\cal C}_{(4)}=q^{-N}([N]-a^+a)\eqno(9a)$$
commutes with all the operators i.e. $a,~a^+,~N$. Thus one can
write
$$a^+a=[N]-q^N{\cal C}_{(4)}.\eqno(9b)$$
Moreover, we also have , using (4)
$$aa^+=qa^+a + q^{-N}=[N+1]-q^{N+1}{\cal C}_{(4)}.\eqno(9c)$$
It is apparent then that (4) implies (2) only if
${\cal C}_{(4)}=0$.
To show inequivalence it is sufficient to show that there exist
representations of (4) in which ${\cal C}_{(4)}$ is non zero.
To this end,  we have for $n=0,1,2...$(see ref. [9])
$$\eqalignno{a^+\vert n> &= q^{-\nu_{0}/2}[n+1]^{1/2}\vert n+1>,
&(10a)\cr
 a\vert n> &= q^{-\nu_{0}/2}[n]^{1/2}\vert n-1>, &(10b)\cr
N\vert n> &= (\nu_{0}+n)\vert n>,&(10c)}$$
which suitably represents (4) and (6). Note that this
representation carries a free parameter $\nu_0$ and is more general
then (3) above. In this representation one has
$${\cal C}_{(4)}\vert n>=q^{-\nu_0}[\nu_0]\vert n>\eqno(11)$$
which shows that for $\nu_0\ne 0$, ${\cal C}_{(4)}$ cannot be
regarded as the null operator. Thus we can surmise that (4), in
general, do not imply (2).
\par \noindent
$\underline {\rm Case~(ii):~Between~(2)~and~(5).}$
\par \noindent
Using arguments paralleling above, it is easy to show that (2) implies
(5) but not the converse. Here ${\cal C}_{(5)}$ is similar to
${\cal C}_{(4)}$ with $q\leftrightarrow q^{-1}$.
\eject
\par \noindent
$\underline {\rm Case~(iii):~Between~(2)~and~(7).}$
\par \noindent
It is obvious that (2) implies (7). For the converse we construct
the Casmir operator ${\cal C}_{(7)}$ for (7) which reads as
$${\cal C}_{(7)}=[N]-a^+a\eqno(12a)$$
or
$$a^+a=[N]-{\cal C}_{(7)}.\eqno(12b)$$
Using (7), we also have
$$aa^+=[N+1]-{\cal C}_{(7)}.\eqno(12c)$$
{}From these it is clear that (7) implies (2) only if
${\cal C}_{(7)}=0$.
Again for non-equivalence,
it suffices to show that ${\cal C}_{(7)}$ is not zero in some
representation. In this case one can construct the following
representation:
$$\eqalignno{a^+\vert n> &= ([n+1-\nu_0]+[\nu_0])^{1/2}\vert n+1>,
&(13a)\cr
 a\vert n> &= ([n-\nu_0]+[\nu_0])^{1/2}\vert n-1>, &(13b)\cr
N\vert n> &= (n-\nu_{0})\vert n>,&(13c)}$$
for $n=0,1,2...$ in which
$${\cal C}_{(7)}\vert n> = -[\nu_0]\vert n>.\eqno(14)$$
It is evident then that ${\cal C}_{(7)}\ne 0$ for $\nu_0\ne 0$.
\par \noindent
$\underline {\rm Case~(iv):~Between~(4)~and~(5).}$
\par \noindent
{}From (9b) and (9c) we have
$$\eqalignno{aa^+-q^{-1}a^+a &=[N+1]-q^{N+1}{\cal C}_{(4)}-q^{-1}
([N]-q^{N}{\cal C}_{(4)})\cr
&= q^N-(q-q^{-1})q^N{\cal C}_{(4)}&(15)}$$
which shows that (4) is not equivalent to (5) since ${\cal
C}_{(4)}\ne 0$ in general. Similarly (5) does not imply (4).
\par \noindent
$\underline {\rm Case~(v):~Between~(4)~and~(7).}$
\par \noindent
{}From (9b) and (9c) we obtain
$$aa^+-a^+a=[N+1]-[N]-(q-1)q^N{\cal C}_{(4)}\eqno(16)$$
which means that (4) does not imply (7). Conversely from (12b) and
(12c) we have
$$\eqalignno{aa^+-qa^+a&=[N+1]-q[N]-{\cal C}_{(7)}+
q{\cal C}_{(7)}\cr
&= q^{-N}+(q-1){\cal C}_{(7)}&(17)}$$
which again demonstrates the inequivalence.
\par \noindent
$\underline {\rm Case~(vi):~Between~(5)~and~(7).}$
\par \noindent
Arguments and conclusions similar to case (v) with
$q\leftrightarrow q^{-1}$.
\par
{}From the results above we can surmise that although the various
forms are interchangeable
in the representation (3), they are nevertheless inequivalent at
the algebraic level. This distinction becomes particularly
important when one deals with pure algebraic constructs. For
instance, when we are considering the Hopf structure of the
$q$-oscillators, it is necessary to distinguish the relation in
question.
Now, due to the equivalence of (2), (4), (5) and (7) {\it in the
representation} (3), it is sometimes implied that they are all Hopf
algebras. Indeed, it has been claimed (see ref. [10]) that
relations (4) {\it together with} (5) which imply (2) and hence
(7) have a Hopf structure defined by
\footnote\dag {In ref. [10] $q^{1/2}$ instead of $q$ is used.}:
$$\eqalignno{
\Delta(a^+) &= (a^+\otimes q^{1/2(N+1/2)}
+ iq^{-1/2(N+1/2)}\otimes a^+)e^{-i\theta/2}, &(18a)\cr
\Delta(a) &= (a\otimes q^{1/2(N+1/2)}
+ iq^{-1/2(N+1/2)}\otimes a)e^{-i\theta/2}, &(18b)\cr
\Delta(N) &= N\otimes {\bf 1}
+ {\bf 1}\otimes N +\gamma {\bf 1}\otimes {\bf 1}, &(18c)\cr
\Delta({\bf 1})&={\bf 1}\otimes {\bf 1},&(18d)\cr
\epsilon (a^+) &=\epsilon (a) = 0, &(18e)\cr
\epsilon (N)&= -\gamma,~~~~~~~~~~~~~~\epsilon ({\bf 1})=1,&(18f)\cr
S(a^+)&= -q^{1/2}a^+,~~~~~~~S(a)=-q^{-1/2}a,&(18g)\cr
S(N)&=-N-2\gamma {\bf 1},~~~~S({\bf 1})={\bf 1},&(18h)\cr}$$
where $\gamma = {{1}\over{2}}-{{i\theta}\over{{\rm ln}~q}}$ and
$\theta=(\pi/2) + 2\pi l$, $l\in {\bf Z}$. Here
the maps $\Delta$,
$\epsilon$, $S$ which are the coproduct, counit and antipode
respectively, satisfy
$$ \Delta (ab)=\Delta (a)\Delta (b),~~~~\epsilon (ab)
=\epsilon (a)\epsilon(b),~~~~S(ab)=S(b)S(a)\eqno(19)$$
for any two elements $a$, $b$ of the Hopf algebra. In other words
they are algebra homomorphisms (antihomomorphism for the case of
$S$). Although these constitute a Hopf structure for relation (7)
together with (6), they do not for relations (4), (5), and (2).
To see why this is so, consider relation (4) as an example. By
applying $\Delta$ on both sides we have using (19),
$$\eqalignno{
\Delta(a)\Delta(a^+)-q\Delta(a^+)\Delta(a)
&= -i\{q^{-N}\otimes q^{N+1/2}
+i(1-q)q^{-1/2}q^{-1/2N}a\otimes q^{1/2N}a^+\cr
& -q^{-(N+1/2)}\otimes q^{-N}
+i(1-q)q^{1/2}q^{-1/2N}a^+\otimes q^{1/2N}a\}
&(20)\cr}$$
whereas
$$\Delta(q^{-N})=iq^{-1/2}q^{-N}\otimes q^{-N}\eqno(21)$$
which means that
$$\Delta(a)\Delta(a^+)-q\Delta(a^+)\Delta(a)\ne\Delta(q^{-N}).\eqno(22)$$
Similarly one can easily verify that $\Delta$ is also incompatible with
(2) and (5). Thus the Hopf structure is valid only for relation (7)
and not for the rest.
\vskip .5cm
\par \noindent
{\bf 3. Generalized $q$-Oscillator Algebra}
\par
Recently some authors [11] have considered a
generalized version of (4),
$$aa^+-qa^+a=q^{\alpha N +\beta}\eqno(23)$$
together with (6)
and attempted to give it a Hopf structure. However, the coproduct
defined there fails the compatibility requirement,
in the same way as (18a) and (18b) failed for relations (4), (5)
and (2).
Specifically, the proposed coproduct,
$$\eqalignno{
\Delta(a^+) &= a^+\otimes q^{1/2(\alpha N+\beta)}
+ q^{-1/2(\alpha N+\beta)}\otimes a^+, &(24a)\cr
\Delta(a) &= a\otimes q^{1/2(\alpha N+\beta)}
+ q^{-1/2(\alpha N+\beta)}\otimes a, &(24b)\cr
\Delta(N) &= N\otimes {\bf 1}
+ {\bf 1}\otimes N +(\beta/\alpha) {\bf 1}\otimes {\bf 1},
&(24c)\cr}$$
fails with respect to (23):
$$\Delta(a)\Delta(a^+)-q\Delta(a^+)\Delta(a)\ne\Delta(q^{\alpha N+\beta}).
\eqno(24)$$
In this section we furnish a generalized version of the $q$-oscillator
algebra which can be endowed with a Hopf structure.
\par
As seen from the previous section, among the various forms of the
$q$-oscillator algebra relation (7) is the only version
that supports a Hopf structure. So it is
conceivable that any generalization would most likely be based on (7)
rather than (4).
To write down a generalized version of (7), it is instructive to
consider again the relationship between (4) and (7). Instead of
using a representation in which ${\cal C}_{(4)}$ is zero (see (16)),
one can also obtain (7) by considering {\it both} (4) {\it and} (5).
When taken together they imply (2) which in turn imply (7). It is
worth noting that (5) is the $q\leftrightarrow q^{-1}$ analogue of
(4). Now let
us apply this procedure to (23). Since $\alpha$ and $\beta$ are arbitrary
we can replace $\alpha \to -\alpha$ and $\beta \to -\beta$ and rewrite
(23) as
$$aa^+-qa^+a=q^{-\alpha N -\beta}.\eqno(26)$$
Its $~q\leftrightarrow q^{-1}$ analogue is given by
$$aa^+-q^{-1}a^+a=q^{\alpha N +\beta}\eqno(27)$$
which together with (26) imply that
$$\eqalignno{
&aa^+ = [\alpha N+\beta], &(28a)\cr
&a^+a = [\alpha N+\beta +1]. &(28b)\cr}$$
This in turn leads us to
$$[a,a^+]=[\alpha N +\beta +1]-[\alpha N +\beta]\eqno(29)$$
or
$$[a,a^+]=[\alpha N +\beta_{1}]-[\alpha N +\beta_{2}]\eqno(30a)$$
where
$$\beta_1-\beta_2=1,\eqno(30b)$$
which is a generalization of (7) we sought. To be as general as possible we
will also modify (6) somewhat by introducing a free parameter
$\eta$ into the commutation relations:
$$[N,a^+]=\eta a^+,~~~~~~[N,a]=-\eta a. \eqno(31)$$
It is worth noting that the algebra  composed of (30) and (31) admits a
non-trivial central term which is given by
$${\cal C}=a^+a - {{1}\over{2}}
{{{\rm sinh}(\epsilon (\alpha N +\beta +1/2-\alpha\eta /2))}
\over {{\rm cosh}(\epsilon/2){\rm sinh}(\epsilon \alpha\eta /2)}}.
\eqno(32)$$
\par
Now it is also important to note that, as in the case between (2) and
(7), relations (30) do not necessarily imply (28) although the
latter have been used in constructing the former. This can be
demonstrated easily by constructing a representation of (30) and
(31) in which (26), (27) and (28) do not hold individually. Indeed,
in the basis $\{\vert n>\}$, $n=0,1,2,...$,
a representation of (30) and (31) is given  by
$$\eqalignno{
a\vert n> &= \{ {{{\rm cosh}(\epsilon \alpha
(\nu_0+(n-1)\eta/2)+\epsilon (\beta +1/2))
{\rm sinh}(\epsilon \eta \alpha n/2)}
\over{{\rm cosh}(\epsilon /2){\rm sinh}(\epsilon \eta\alpha/2)}}\}^{1/2}
\vert n-1 >, &(33a)\cr
a^+\vert n> &= \{ {{{\rm cosh}(\epsilon \alpha
(\nu_0+n\eta/2)+\epsilon (\beta +1/2))
{\rm sinh}(\epsilon \eta \alpha (n+1)/2)}
\over{{\rm cosh}(\epsilon /2){\rm sinh}(\epsilon \eta\alpha/2)}}\}^{1/2}
\vert n+1 >, &(33b)\cr
N\vert n> &= (\nu_0 +n\eta)\vert n>,&(33c)\cr}$$
where we have taken $\epsilon = {\ln}~q$ and
$\nu_0$ is a free parameter which characterizes the representation.
In this representation, it not difficult to see the inequivalence between
(30) and any one among (26)-(28). This is true even when $\eta$ is
set to 1.
\par
Now let us turn to the Hopf structure associated with (30) and (31).
We start by considering the
associative algebra
$\cal H$ generated by $\{ {\bf 1},a^+,a,N\}$
and postulating the following for the coproduct,
counit and antipode:
$$\eqalignno{
\Delta(a^+) &= c_{1}a^+\otimes q^{\alpha_1 N}
+ c_{2}q^{\alpha_2 N}\otimes a^+, &(34a)\cr
\Delta(a) &= c_{3}a\otimes q^{\alpha_3 N}
+ c_{4}q^{\alpha_4 N}\otimes a, &(34b)\cr
\Delta(N) &= c_{5}N\otimes {\bf 1}
+ c_{6}{\bf 1}\otimes N +\gamma {\bf 1}\otimes {\bf 1}, &(34c)\cr
\Delta({\bf 1})&={\bf 1}\otimes {\bf 1},&(34d)\cr
\epsilon (a^+) &=c_{7}~~~~~~~~~~~~~~~\epsilon (a) = c_{8}, &(34e)\cr
\epsilon (N)&= c_{9},~~~~~~~~~~~~~~~~\epsilon ({\bf 1})=1,&(34f)\cr
S(a^+)&= -c_{10}a^+,~~~~~~~S(a)=-c_{11}a,&(34g)\cr
S(N)&=-c_{12}N+c_{13} {\bf 1},~~~~S({\bf 1})={\bf 1}.&(34h)\cr}$$
Here $c_{i},~i=1,2,...13$, $\alpha_{i},~i=1,2,...4$ and $\gamma$
are constants to be determined. These constants are obtained by
requiring that $\Delta$, $\epsilon$ and $S$ satisfy the
coassociativity, counit and antipode axioms respectively:
$$\eqalignno{&({\rm id}\otimes \Delta)\Delta(h)=
(\Delta \otimes {\rm id})\Delta(h), &(35a)\cr
&({\rm id}\otimes \epsilon)\Delta(h)=
(\epsilon \otimes {\rm id})\Delta(h)=h, &(35b)\cr
&m({\rm id}\otimes S)\Delta(h)=
m(S \otimes {\rm id})\Delta(h)=\epsilon(h){\bf 1}, &(35c)\cr}$$
where $h\in {\cal H}$ and $m:{\cal H}\otimes {\cal H}\to {\cal H}$
is the multiplication map. By substituting the different generators
of $\cal H$ into (35) and noting that
$$\eqalignno{a^+q^{\alpha N}&= q^{-\alpha \eta}q^{\alpha N}a^+,&(36a)\cr
aq^{\alpha N}&= q^{\alpha \eta}q^{\alpha N}a,&(36b)\cr}$$
for an arbitrary $\alpha$, we obtain
$$\eqalignno{
&c_{1}=q^{\alpha_1\gamma},~~~~c_{2}=q^{\alpha_2\gamma},~~~~
c_{3}=q^{\alpha_3\gamma},~~~~c_{4}=q^{\alpha_4\gamma}~~~~c_{5}=1,\cr
&c_{6}=1,~~~~~~~~c_{7}=0,~~~~~~~~c_{8}=0,~~~~~~~c_{9}=-\gamma,
{}~~~~~~c_{10}=q^{\alpha_1\eta},\cr
&c_{11}=q^{-\alpha_3\eta},~~~~c_{12}=1,~~~~~~
c_{13}=-2\gamma,~~~~\alpha_2=-\alpha_1,~~~~\alpha_4=-\alpha_3,&(37)}$$
which essentially fixes 15 of the 18 constants in (34).
We must also require that $\Delta$, $\epsilon$ and $S$ be algebra
homomorphisms. Here further constraints arise when we set
$$\Delta (a)\Delta (a^+)-\Delta (a^+)\Delta (a)=\Delta ([\alpha N+\beta_1]-
[\alpha N+\beta_2]).\eqno(38)$$
For this to be satisfied we must impose the following:
$$\eqalignno{
&q^{(\alpha_1-\alpha_3)\eta}=1,&(39a)\cr
&\alpha_1 + \alpha_3 =\alpha,&(39b)\cr
&q^{2\alpha \gamma}= -q^{\beta_1 + \beta_2}.&(39c)\cr}$$
With these, the homomorphisms $\epsilon$ and $S$ entail
no further constraints. For real $q$, eqns. (39) imply that
$$\alpha_1=\alpha_3={{1}\over {2}} \alpha\eqno(40a)$$
and
$$\gamma = {{\beta_1+\beta_2}\over {2\alpha}}-{{i(2k+1)\pi}\over {2\alpha
{}~{\rm ln}~q}}~~~~~~k\in {\bf Z}\eqno(40b)$$
which now fixes all the constants in (34).
\par
To summarize briefly, the Hopf structure for $\cal H$ with defining
relations (30) and (31) reads as
$$\eqalignno{
\Delta(a^+) &= a^+\otimes q^{{{1}\over {2}}\alpha(N+\gamma)}
+ q^{-{{1}\over {2}}\alpha(N+\gamma)}\otimes a^+, &(41a)\cr
\Delta(a) &= a \otimes q^{{{1}\over {2}}\alpha(N+\gamma)}
+ q^{-{{1}\over {2}}\alpha(N+\gamma)}\otimes a, &(41b)\cr
\Delta(N) &= N\otimes {\bf 1}
+ {\bf 1}\otimes N +\gamma {\bf 1}\otimes {\bf 1}, &(41c)\cr
\Delta({\bf 1})&={\bf 1}\otimes {\bf 1},&(41d)\cr
\epsilon (a^+) &=\epsilon (a) = 0, &(41e)\cr
\epsilon (N)&= -\gamma,~~~~~~~~~~~~~~\epsilon ({\bf 1})=1,&(41f)\cr
S(a^+)&= -q^{{{1}\over {2}}\alpha \eta}a^+,~~~~~~~
S(a)=-q^{-{{1}\over {2}}\alpha \eta}a,&(41g)\cr
S(N)&=-N-2\gamma {\bf 1},~~~~S({\bf 1})={\bf 1},&(41h)\cr}$$
where $\gamma$ satisfies (40b).
Note that by setting $\eta=1$, $\beta_1=1$,
$\beta_2=0$, $\alpha = 1$ and putting $k=2l,~~l\in {\bf Z}$ in
(40b) we recover the Hopf structure associated with (6) and (7).
\vskip .5cm
\par\noindent
{\bf 4. Multimode $q$-Oscillators }
\par
Various multimode extensions of the $q$-oscillators have been
proposed [4,5,12,13]. In particular the extension of (4) {\it and} (5) or
equivalently that of (2) consist of taking $p$ independent oscillators
(mutually commuting)
$\{a_i,a^+_i,N_i \vert~i=1,2,...p\}$ with the relations [13]
$$\eqalignno{
&a_ia^+_j-(1+\delta_{ij}(q-1))a^+_ja_i=\delta_{ij}q^{-N_i}, &(42a)\cr
&a_ia^+_j-(1+\delta_{ij}(q^{-1}-1))a^+_ja_i=\delta_{ij}q^{N_i}, &(42b)\cr
&[a_i,a_j]=[a_i^+,a_j^+]=0,&(42c)\cr
&[N_i,a_j]=-\delta_{ij}a_j,~~~~~[N_i,a^+_j]=\delta_{ij}a^+_j. &(42d)\cr}$$
\par
Here we present a multimode extension of (30) and show that it also
supports a non-cocommutative Hopf structure. To this end we propose
the following relations for the set of $p$ oscillators:
$$\eqalignno{
&[a_i,a^+_j]=([\alpha_iN_i+\beta_i+1]
-[\alpha_iN_i+\beta_i])\delta_{ij},&(43a)\cr
&[a_i,a_j]=[a_i^+,a_j^+]=0,&(43b)\cr
&[N_i,a_j]=-\eta_ia_j\delta_{ij},~~~~~
[N_i,a^+_j]=\eta_ia^+_j\delta_{ij}&(43c)\cr
}$$
where $\alpha_i$, $\beta_i$ and $\eta_i$ ($i=1,2,...p$) are free parameters.
Then it can easily be shown that the associative algebra
generated by $\{{\bf 1},a_i,a^+_i,N_i\}$, $i=1,2,...p$ with the
above defining relations admits the following non-cocommutative
Hopf structure:
$$\eqalignno{
\Delta(a_i^+) &= a_i^+\otimes q^{{{1}\over {2}}\alpha_i(N_i+\gamma_i)}
+ q^{-{{1}\over {2}}\alpha_i(N_i+\gamma_i)}\otimes a_i^+, &(44a)\cr
\Delta(a_i) &= a_i \otimes q^{{{1}\over {2}}\alpha_i(N_i+\gamma_i)}
+ q^{-{{1}\over {2}}\alpha_i(N_i+\gamma_i)}\otimes a_i, &(44b)\cr
\Delta(N_i) &= N_i\otimes {\bf 1}
+ {\bf 1}\otimes N_i +\gamma_i {\bf 1}\otimes {\bf 1}, &(44c)\cr
\Delta({\bf 1})&={\bf 1}\otimes {\bf 1},&(44d)\cr
\epsilon (a_i^+) &=\epsilon (a_i) = 0, &(44e)\cr
\epsilon (N_i)&= -\gamma_i,~~~~~~~~~~~~~~\epsilon ({\bf 1})=1,&(44f)\cr
S(a_i^+)&= -q^{{{1}\over {2}}\alpha_i \eta_i}a_i^+,~~~~~~~
S(a_i)=-q^{-{{1}\over {2}}\alpha_i \eta_i}a_i,&(44g)\cr
S(N_i)&=-N_i-2\gamma_i {\bf 1},~~~~S({\bf 1})={\bf 1},&(44h)\cr}$$
with
$$\gamma_i = {{2\beta_i+1}\over {2\alpha_i}}
-{{i(2k_i+1)\pi}\over {2\alpha_i
{}~{\rm ln}~q}}~~~~k_i\in {\bf Z}.\eqno(45)$$
In verifying the homomorphism property we have used
$$\eqalignno{a_i^+q^{\rho N_j}&= q^{-\rho \eta_j\delta_{ij}}
q^{\rho N_j}a_i^+,&(46a)\cr
a_iq^{\rho N_j}&= q^{\rho \eta_j\delta_{ij}}q^{\rho N_j}a_i,&(46b)\cr}$$
for an arbitrary $\rho$. Here the Hopf structure is essentially the
Hopf structure of each oscillator taken independently. It is
interesting to note that the oscillators can also be coupled in
a non-trivial way. To see how this can be accomplished, let us
examine relations (39) closely. If we allow $\eta$ to be
complex and relate it to $q$ via
$$\eta={{2\pi i}\over {{\rm ln}~q}}\eqno(47)$$
then relation (39a) implies that $\alpha_1-\alpha_3 = l$, ($l\in
{\bf Z}$). Now this means that we can assign integer values to
$\alpha_1$ and $\alpha_3$ which in turn allows the indexing of
oscillators. For instance, if we set $\alpha_1=m$ and $\alpha_3=n$
then the oscillators can be indexed as $a_m$ and $a^+_n$
respectively. This effectively permits a number of oscillators to be
considered together. Moreover, with $\alpha$ also being integer
valued, as a consequence of (39b), the commutation relations
between the various oscillators become non trivial. As for (39c), we have
$$\beta_1+\beta_2=2(\alpha_1+\alpha_3)\gamma+i{{(2k+1)\pi}\over{{\rm
ln}~q}}~~~~~~~k\in {\bf Z}.\eqno(48)$$
Then by putting $k=0$ (for simplicity) and using (30b), we obtain
$$\eqalignno{
\beta_1 &=(\alpha_1+\alpha_3)\gamma
+ {{i\pi}\over {2~{\rm ln}~q}}+{{1}\over {2}},&(49a)\cr
\beta_2 &=(\alpha_1+\alpha_3)\gamma
+ {{i\pi}\over {2~{\rm ln}~q}}-{{1}\over {2}}.&(49b)\cr}$$
With these, we have
\footnote\dag {Here we have set $\alpha_1=m$ and $\alpha_3=n$ and
the corresponding oscillators by $a_m$ and $a^+_n$ respectively.}
$$\eqalignno{
[a_m,a^+_n] &= [(m+n)(N+\gamma)+{{i\pi}\over {2~{\rm ln}~q}}
+{{1}\over {2}}]-
[(m+n)(N+\gamma)+{{i\pi}\over {2~{\rm ln}~q}}-{{1}\over {2}}]\cr
&= i{{{\rm sinh}(\epsilon
(m+n)(N+\gamma))}\over {{\rm cosh}(\epsilon/2)}}&(50)\cr}$$
where we have taken $q=e^{\epsilon}$. Thus the commutation
relations for a system of $p$ oscillators can be written as
$$\eqalignno{
&[a_m,a^+_n]
= i{{{\rm sinh}(\epsilon (m+n)(N+\gamma))}\over {{\rm cosh}(\epsilon/2)}}
&(51a)\cr
&[a_m,a_n]=[a_m^+,a_n^+]=0,&(51b)\cr
&[N,a_m]=-{{2\pi i}\over {{\rm ln}~q}} a_m,~~~~~
[N,a^+_m]= {{2\pi i}\over {{\rm ln}~q}}a^+_m &(51c)\cr}$$
where $m,n=1,2,...p$. It is important to note that unlike the
previous case we have only one $N$ operator. The corresponding Hopf
structure  is then given by
$$\eqalignno{
\Delta(a_m^+) &= a_m^+\otimes q^{m(N+\gamma)}
+ q^{-m(N+\gamma)}\otimes a_m^+, &(52a)\cr
\Delta(a_m) &= a_m\otimes q^{m(N+\gamma)}
+ q^{-m(N+\gamma)}\otimes a_m, &(52b)\cr
\Delta(N) &= N\otimes {\bf 1}
+ {\bf 1}\otimes N +\gamma {\bf 1}\otimes {\bf 1}, &(52c)\cr
\Delta({\bf 1})&={\bf 1}\otimes {\bf 1},&(52d)\cr
\epsilon (a_m^+) &=\epsilon (a_m) = 0, &(52e)\cr
\epsilon (N)&= -\gamma,~~~~~~~~~~~~~~\epsilon ({\bf 1})=1,&(52f)\cr
S(a_m^+)&= -a_m^+,~~~~~~~~~~~~
S(a_m)=-a_m,&(52g)\cr
S(N)&=-N-2\gamma {\bf 1},~~~~S({\bf 1})={\bf 1}.&(52h)\cr}$$
\vskip .5cm
\par\noindent
{\bf 5. Conclusion}
\par
In this paper we have considered the various forms of the
$q$-oscillator algebra and shown that, contrary to the commonly
held notion, they are actually not equivalent. It is also pointed
out that the Hopf structure found for one of these versions does not
extend to the rest by virtue of this inequivalence. For the algebra
that is a quantum group, we have given its generalization
together with the associated Hopf structure. Based on this
generalization we have also furnished two multimode extensions.
In the first example, we have considered a set of (mutually
commuting) independent oscillators and shown that the Hopf
structure of each oscillator system extends naturally to the
multimode case. For the second example, we have presented a Hopf
algebra comprising of a set of non-commuting oscillators.

\vfill\eject
\noindent
{\bf REFERENCES}
\ref 1 Faddev L D 1982 {\it Les Houches Lectures}
(Elsevier,Amsterdam, 1984)
\spa {~} Kulish P P and Sklyanin E K  1982 {\it Lecture Notes in
Physics Vol.151} (Springer, Berlin);
\ref 2 Drinfeld V G 1986 {\it Proc. Intern. Congress of
Mathematicians} (Berkley) Vol. 1 pg 798;
\ref 3 Sklyanin E K 1982, {\it Funct. Anal. Appl.} {\bf 16} 262;
\ref 4 Kulish P P and Reshetikhin N Y (1983) {\it J. Sov. Math.}
{\bf 23} 2435;
\ref 5 Macfarlane A J 1989 {\it J. Phys.} {\bf A22} 4581;
\ref 6 Biedenharn L C 1989 {\it J. Phys.} {\bf A22} L873;
\ref 7 Ng Y J 1990 {\it J. Phys.} {\bf A23} 1023;
\ref 8 Hong Yan 1990 {\it J. Phys.} {\bf A23} L1155;
\ref 9 Rideau G 1992 {\it Lett. Math. Phys.} {\bf 24} 147;
\ref {10} Floreanini R and Vinet L 1991 {\it Lett. Math. Phys.}
{\bf 22} 45;
\ref {11} Chung W, Chung K, Nam S and Um C 1993 {\it Phys. Lett.}
{\bf A 183} 363;
\ref {12} Pusz W and Woronowicz S L (1989) {\it Rep. Math. Phys.}
{\bf 27} 231;
\ref {13} Hayashi T (1990) {\it Commun. Math. Phys.} {\bf 127} 129;
\vfill
\bye